\documentclass[conference]{IEEEtran}

\makeatletter

\newcommand{\Rmnum}[1]{\expandafter\@slowromancap\romannumeral #1@}
\makeatother
\usepackage{array}
\usepackage{multirow}
\usepackage{diagbox}
\usepackage{amsthm,amsmath,amsfonts,syntonly}
\usepackage{amssymb}
\usepackage{epsfig} 
\usepackage{epstopdf}
\usepackage{epsf,graphics,graphicx}
\usepackage{cite,bm,color,url,textcomp}
\usepackage{float}
\usepackage{algorithm}
\usepackage{algorithm}
\usepackage{algorithmicx}
\usepackage{algpseudocode}
\usepackage{mathrsfs}
\usepackage{arydshln}
\usepackage{stfloats}
\newcommand\blfootnote[1]{%
  \begingroup
  \renewcommand\thefootnote{}\footnote{#1}%
  \addtocounter{footnote}{-1}%
  \endgroup
}
\algdef{SE}[DOWHILE]{Do}{doWhile}{\algorithmicdo}[1]{\algorithmicwhile\ #1}%

\newcommand{\g}[1]{\bm #1}

\begin{document}
\newcommand{\fs}{\hspace{0.07in}}
\newcommand{\bs}{\hspace{-0.1in}}
\newcommand{\re}{{\rm Re} \, }
\newcommand{\e}{{\rm E} \, }
\newcommand{\p}{{\rm P} \, }
\newcommand{\cn}{{\cal CN} \, }
\newcommand{\n}{{\cal N} \, }
\newcommand{\ba}{\begin{array}}
\newcommand{\ea}{\end{array}}
\newcommand{\be}{\begin{displaymath}}
\newcommand{\ee}{\end{displaymath}}
\newcommand{\ben}{\begin{equation}}
\newcommand{\een}{\end{equation}}
\newcommand{\bena}{\begin{eqnarray}}
\newcommand{\eena}{\end{eqnarray}}
\newcommand{\beqa}{\begin{eqnarray*}}
\newcommand{\enqa}{\end{eqnarray*}}
\newcommand{\f}{\frac}
\newcommand{\bc}{\begin{center}}
\newcommand{\ec}{\end{center}}
\newcommand{\bi}{\begin{itemize}}
\newcommand{\ei}{\end{itemize}}
\newcommand{\benu}{\begin{enumerate}}
\newcommand{\eenu}{\end{enumerate}}
\newcommand{\bdes}{\begin{description}}
\newcommand{\edes}{\end{description}}
\newcommand{\bt}{\begin{tabular}}
\newcommand{\et}{\end{tabular}}
\newcommand{\vs}{\vspace}
\newcommand{\hs}{\hspace}
\newcommand{\sort}{\rm sort \,}

\newcommand \thetabf{{\mbox{\boldmath$\theta$\unboldmath}}}
\newcommand{\Phibf}{\mbox{${\bf \Phi}$}}
\newcommand{\Psibf}{\mbox{${\bf \Psi}$}}
\newcommand \alphabf{\mbox{\boldmath$\alpha$\unboldmath}}
\newcommand \betabf{\mbox{\boldmath$\beta$\unboldmath}}
\newcommand \gammabf{\mbox{\boldmath$\gamma$\unboldmath}}
\newcommand \deltabf{\mbox{\boldmath$\delta$\unboldmath}}
\newcommand \epsilonbf{\mbox{\boldmath$\epsilon$\unboldmath}}
\newcommand \zetabf{\mbox{\boldmath$\zeta$\unboldmath}}
\newcommand \etabf{\mbox{\boldmath$\eta$\unboldmath}}
\newcommand \iotabf{\mbox{\boldmath$\iota$\unboldmath}}
\newcommand \kappabf{\mbox{\boldmath$\kappa$\unboldmath}}
\newcommand \lambdabf{\mbox{\boldmath$\lambda$\unboldmath}}
\newcommand \mubf{\mbox{\boldmath$\mu$\unboldmath}}
\newcommand \nubf{\mbox{\boldmath$\nu$\unboldmath}}
\newcommand \xibf{\mbox{\boldmath$\xi$\unboldmath}}
\newcommand \pibf{\mbox{\boldmath$\pi$\unboldmath}}
\newcommand \rhobf{\mbox{\boldmath$\rho$\unboldmath}}
\newcommand \sigmabf{\mbox{\boldmath$\sigma$\unboldmath}}
\newcommand \taubf{\mbox{\boldmath$\tau$\unboldmath}}
\newcommand \upsilonbf{\mbox{\boldmath$\upsilon$\unboldmath}}
\newcommand \phibf{\mbox{\boldmath$\phi$\unboldmath}}
\newcommand \varphibf{\mbox{\boldmath$\varphi$\unboldmath}}
\newcommand \chibf{\mbox{\boldmath$\chi$\unboldmath}}
\newcommand \psibf{\mbox{\boldmath$\psi$\unboldmath}}
\newcommand \omegabf{\mbox{\boldmath$\omega$\unboldmath}}
\newcommand \Sigmabf{\hbox{$\bf \Sigma$}}
\newcommand \Upsilonbf{\hbox{$\bf \Upsilon$}}
\newcommand \Omegabf{\hbox{$\bf \Omega$}}
\newcommand \Deltabf{\hbox{$\bf \Delta$}}
\newcommand \Gammabf{\hbox{$\bf \Gamma$}}
\newcommand \Thetabf{\hbox{$\bf \Theta$}}
\newcommand \Lambdabf{\hbox{$\bf \Lambda$}}
\newcommand \Xibf{\hbox{\bf$\Xi$}}
\newcommand \Pibf{\hbox{\bm$\Pi$}}
\newcommand \abf{{\bf a}}
\newcommand \bbf{{\bf b}}
\newcommand \cbf{{\bf c}}
\newcommand \dbf{{\bf d}}
\newcommand \ebf{{\bf e}}
\newcommand \fbf{{\bf f}}
\newcommand \gbf{{\bf g}}
\newcommand \hbf{{\bf h}}
\newcommand \ibf{{\bf i}}
\newcommand \jbf{{\bf j}}
\newcommand \kbf{{\bf k}}
\newcommand \lbf{{\bf l}}
\newcommand \mbf{{\bf m}}
\newcommand \nbf{{\bf n}}
\newcommand \obf{{\bf o}}
\newcommand \pbf{{\bf p}}
\newcommand \qbf{{\bf q}}
\newcommand \rbf{{\bf r}}
\newcommand \sbf{{\bf s}}
\newcommand \tbf{{\bf t}}
\newcommand \ubf{{\bf u}}
\newcommand \vbf{{\bf v}}
\newcommand \wbf{{\bf w}}
\newcommand \xbf{{\bf x}}
\newcommand \ybf{{\bf y}}
\newcommand \zbf{{\bf z}}
\newcommand \rbfa{{\bf r}}
\newcommand \xbfa{{\bf x}}
\newcommand \ybfa{{\bf y}}
\newcommand \Abf{{\bf A}}
\newcommand \Bbf{{\bf B}}
\newcommand \Cbf{{\bf C}}
\newcommand \Dbf{{\bf D}}
\newcommand \Ebf{{\bf E}}
\newcommand \Fbf{{\bf F}}
\newcommand \Gbf{{\bf G}}
\newcommand \Hbf{{\bf H}}
\newcommand \Ibf{{\bf I}}
\newcommand \Jbf{{\bf J}}
\newcommand \Kbf{{\bf K}}
\newcommand \Lbf{{\bf L}}
\newcommand \Mbf{{\bf M}}
\newcommand \Nbf{{\bf N}}
\newcommand \Obf{{\bf O}}
\newcommand \Pbf{{\bf P}}
\newcommand \Qbf{{\bf Q}}
\newcommand \Rbf{{\bf R}}
\newcommand \Sbf{{\bf S}}
\newcommand \Tbf{{\bf T}}
\newcommand \Ubf{{\bf U}}
\newcommand \Vbf{{\bf V}}
\newcommand \Wbf{{\bf W}}
\newcommand \Xbf{{\bf X}}
\newcommand \Ybf{{\bf Y}}
\newcommand \Zbf{{\bf Z}}
\newcommand \Omegabbf{{\bf \Omega}}
\newcommand \Rssbf{{\bf R_{ss}}}
\newcommand \Ryybf{{\bf R_{yy}}}
\newcommand \Cset{{\cal C}}
\newcommand \Rset{{\cal R}}
\newcommand \Zset{{\cal Z}}
\newcommand{\otheta}{\stackrel{\circ}{\theta}}
\newcommand{\defeq}{\stackrel{\bigtriangleup}{=}}
\newcommand{\oabf}{{\bf \breve{a}}}
\newcommand{\odbf}{{\bf \breve{d}}}
\newcommand{\oDbf}{{\bf \breve{D}}}
\newcommand{\oAbf}{{\bf \breve{A}}}
\renewcommand \vec{{\mbox{vec}}}
\newcommand{\Acalbf}{\bf {\cal A}}
\newcommand{\calZbf}{\mbox{\boldmath $\cal Z$}}
\newcommand{\feop}{\hfill \rule{2mm}{2mm} \\}
\newtheorem{theorem}{Theorem}[section]

\newcommand{\Rnum}{{\mathbb R}}
\newcommand{\Cnum}{{\mathbb C}}
\newcommand{\Znum}{{\mathbb Z}}
\newcommand{\Enum}{{\mathbb E}}

\newcommand{\Ical}{{\cal I}}
\newcommand{\Mcal}{{\cal M}}
\newcommand{\Pcal}{{\cal P}}
\newcommand{\Ccal}{{\cal C}}
\newcommand{\Dcal}{{\cal D}}
\newcommand{\Hcal}{{\cal H}}
\newcommand{\Ocal}{{\cal O}}
\newcommand{\Rcal}{{\cal R}}
\newcommand{\Zcal}{{\cal Z}}
\newcommand{\Xcal}{{\cal X}}
\newcommand{\zzbf}{{\bf 0}}
\newcommand{\zebf}{{\bf 0}}

\newcommand{\eop}{\hfill $\Box$}

\newcommand{\gss}{\mathop{}\limits}
\newcommand{\gs}{\mathop{\gss_<^>}\limits}

\newcommand{\circlambda}{\mbox{$\Lambda$
             \kern-.85em\raise1.5ex
             \hbox{$\scriptstyle{\circ}$}}\,}

\newcommand{\tr}{\mathop{\rm tr}}
\newcommand{\var}{\mathop{\rm var}}
\newcommand{\cov}{\mathop{\rm cov}}
\newcommand{\diag}{\mathop{\rm diag}}
\def\rank{\mathop{\rm rank}\nolimits}
\newcommand{\ra}{\rightarrow}
\newcommand{\ul}{\underline}
\def\Pr{\mathop{\rm Pr}}
\def\Re{\mathop{\rm Re}}
\def\Im{\mathop{\rm Im}}

\def\submbox#1{_{\mbox{\footnotesize #1}}}
\def\supmbox#1{^{\mbox{\footnotesize #1}}}

%
\newtheorem{Theorem}{Theorem}[section]
\newtheorem{Definition}[Theorem]{Definition}
\newtheorem{Proposition}[Theorem]{Proposition}
\newtheorem{Lemma}[Theorem]{Lemma}
\newtheorem{Corollary}[Theorem]{Corollary}
%
%
\newcommand{\ThmRef}[1]{\ref{thm:#1}}
\newcommand{\ThmLabel}[1]{\label{thm:#1}}
\newcommand{\DefRef}[1]{\ref{def:#1}}
\newcommand{\DefLabel}[1]{\label{def:#1}}
\newcommand{\PropRef}[1]{\ref{prop:#1}}
\newcommand{\PropLabel}[1]{\label{prop:#1}}
\newcommand{\LemRef}[1]{\ref{lem:#1}}
\newcommand{\LemLabel}[1]{\label{lem:#1}}
%

\newcommand \bbs{{\boldsymbol b}}
\newcommand \cbs{{\boldsymbol c}}
\newcommand \dbs{{\boldsymbol d}}
\newcommand \ebs{{\boldsymbol e}}
\newcommand \fbs{{\boldsymbol f}}
\newcommand \gbs{{\boldsymbol g}}
\newcommand \hbs{{\boldsymbol h}}
\newcommand \ibs{{\boldsymbol i}}
\newcommand \jbs{{\boldsymbol j}}
\newcommand \kbs{{\boldsymbol k}}
\newcommand \lbs{{\boldsymbol l}}
\newcommand \mbs{{\boldsymbol m}}
\newcommand \nbs{{\boldsymbol n}}
\newcommand \obs{{\boldsymbol o}}
\newcommand \pbs{{\boldsymbol p}}
\newcommand \qbs{{\boldsymbol q}}
\newcommand \rbs{{\boldsymbol r}}
\newcommand \sbs{{\boldsymbol s}}
\newcommand \tbs{{\boldsymbol t}}
\newcommand \ubs{{\boldsymbol u}}
\newcommand \vbs{{\boldsymbol v}}
\newcommand \wbs{{\boldsymbol w}}
\newcommand \xbs{{\boldsymbol x}}
\newcommand \ybs{{\boldsymbol y}}
\newcommand \zbs{{\boldsymbol z}}

\newcommand \Bbs{{\boldsymbol B}}
\newcommand \Cbs{{\boldsymbol C}}
\newcommand \Dbs{{\boldsymbol D}}
\newcommand \Ebs{{\boldsymbol E}}
\newcommand \Fbs{{\boldsymbol F}}
\newcommand \Gbs{{\boldsymbol G}}
\newcommand \Hbs{{\boldsymbol H}}
\newcommand \Ibs{{\boldsymbol I}}
\newcommand \Jbs{{\boldsymbol J}}
\newcommand \Kbs{{\boldsymbol K}}
\newcommand \Lbs{{\boldsymbol L}}
\newcommand \Mbs{{\boldsymbol M}}
\newcommand \Nbs{{\boldsymbol N}}
\newcommand \Obs{{\boldsymbol O}}
\newcommand \Pbs{{\boldsymbol P}}
\newcommand \Qbs{{\boldsymbol Q}}
\newcommand \Rbs{{\boldsymbol R}}
\newcommand \Sbs{{\boldsymbol S}}
\newcommand \Tbs{{\boldsymbol T}}
\newcommand \Ubs{{\boldsymbol U}}
\newcommand \Vbs{{\boldsymbol V}}
\newcommand \Wbs{{\boldsymbol W}}
\newcommand \Xbs{{\boldsymbol X}}
\newcommand \Ybs{{\boldsymbol Y}}
\newcommand \Zbs{{\boldsymbol Z}}

\newcommand \Absolute[1]{\left\lvert #1 \right\rvert}

\title{Phases Calibration of RIS Using Backpropagation Algorithm}
\author{Wei Zhang$^1$, Bin Zhou$^1$, Tianyi Zhang$^2$, Yi Jiang$^3$, Zhiyong Bu$^1$ \\
$^1$Shanghai Institute of Microsystem and Information Technology, Chinese Academy of Sciences, Shanghai, China\\
Email: {\{wzhang, bin.zhou, zhiyong.bu \}@mail.sim.ac.cn} \\
$^2$Shanghai Radio Equipment Research Institute, Shanghai, China \\
Email: {hm{\_}zty@outlook.com} \\
$^3$School of Information Science and Technology, Fudan University, Shanghai, China \\
Email: {yijiang@fudan.edu.cn}
}


\maketitle
\blfootnote{
Work in this paper was supported by Shanghai Post-doctoral Excellence Program Grant No. 2023689.
The MATLAB codes are available: https://github.com/henryforzhang/qnn-ris-calibration.
}
\begin{abstract}
Reconfigurable intelligent surface (RIS) technology has emerged in recent years as a promising solution to the ever-increasing demand for wireless communication capacity. In practice, however, elements of RIS may suffer from phase deviations, which need to be properly estimated and calibrated. This paper models the problem of over-the-air (OTA) estimation of the RIS elements as a quasi-neural network (QNN) so that the  phase estimates can be obtained using the classic backpropagation (BP) algorithm. We also derive the Cram\'{e}r Rao Bounds (CRBs) for the phases of the RIS elements as a benchmark of the proposed approach. The simulation results verify the effectiveness of the proposed algorithm by showing that the root mean square errors (RMSEs) of the phase estimates are close to the CRBs.
\end{abstract}

%
\IEEEpeerreviewmaketitle

\section{Introduction}
In recent years, the reconfigurable intelligent surface (RIS) technology has been intensively researched for wireless communication, as it can adjust the phase coefficients of its elements to enhance the wireless link quality
\cite{cuiRIS}\cite{9424177} and to mitigate interferences \cite{9839223}. While most of the existing works assume that the tunable phases of the RIS elements are precisely known {\em a priori}, they are unavoidably subject to deviations owing to manufacturing imperfection and aging, which has been shown in \cite[Fig. 1]{9810495}. In the previous work \cite{RAIBFD}, the sum rate performance of the proposed RIS-assisted in-band full-duplex (RAIBFD) system is sensitive to the phase deviations of the RIS elements \cite[Fig. 9]{RAIBFD}. Hence, estimating the phases before applying the RIS to a practical RAIBFD system is a prerequisite.

As related works, conventional array calibration has been widely investigated for phased array radar systems. The paper \cite{4410650508} proposed a so-called rotating element electric field vector (REV) method, which measures the power of received signal as one of phase shifters changes its phase between $0^\circ$ and $180^\circ$ to calibrate an individual antenna element of the phased array. As an improvement to the REV method, the paper \cite{takemura2000phased} considered the effect of phase shifter's phase deviations for more accurate calibration. To be more time-efficient, the paper \cite{4558321} improved the single-element REV method by switching the phases of several elements simultaneously to measure the array power variations. The measured power variations are expanded into a Fourier series, whose coefficients can be used to obtain the amplitudes and phases of multiple antenna elements.

Calibration of phase shifter is also considered in multi-input multi-output (MIMO) systems. The paper \cite{9056550} investigated the over-the-air (OTA) calibration of a phase shifter network (PSN) for mmWave massive MIMO communications, which assumed that phase deviations of phase shifter are the same across different gears. As an extension to \cite{9056550}, the paper \cite{9810495} proposed an OTA PSN calibration method for hybrid MIMO systems given that phase deviations vary at different gears of a phase shifter. For the RIS elements of finite bit resolution, the phase deviations also need to be estimated and calibrated on different gears. The paper \cite{10023531} proposed an OTA calibration algorithm of an $M_{ris}$-element RIS by iteratively estimating the channel and the phases of RIS, but it needs to calculate the inversion of an $M_{ris}\times M_{ris}$ matrix \cite[(17)]{10023531} and performs the iterative Riemannian conjugate gradient (RCG) algorithm in each iteration, yielding excessive computational complexity as $M_{ris}$ grows larger.

In this paper, we propose to model the OTA phase estimation of the RIS elements into a QNN training problem, where the unknown phases and channel state information (CSI) become the weights of the proposed QNN. Then we utilize the backpropagation (BP) algorithm to train the QNN until the cost function converges and obtain the estimated phases. The simulation results verify the effectiveness of the proposed algorithm by showing that the root mean square errors (RMSEs) of the phase estimates closely approach the Cram\'{e}r Rao Bounds (CRBs).


\section{Signal Model and Problem Formulation} \label{SEC2}
\subsection{Signal Model}
\begin{figure}[htb]
\centering
{\psfig{figure=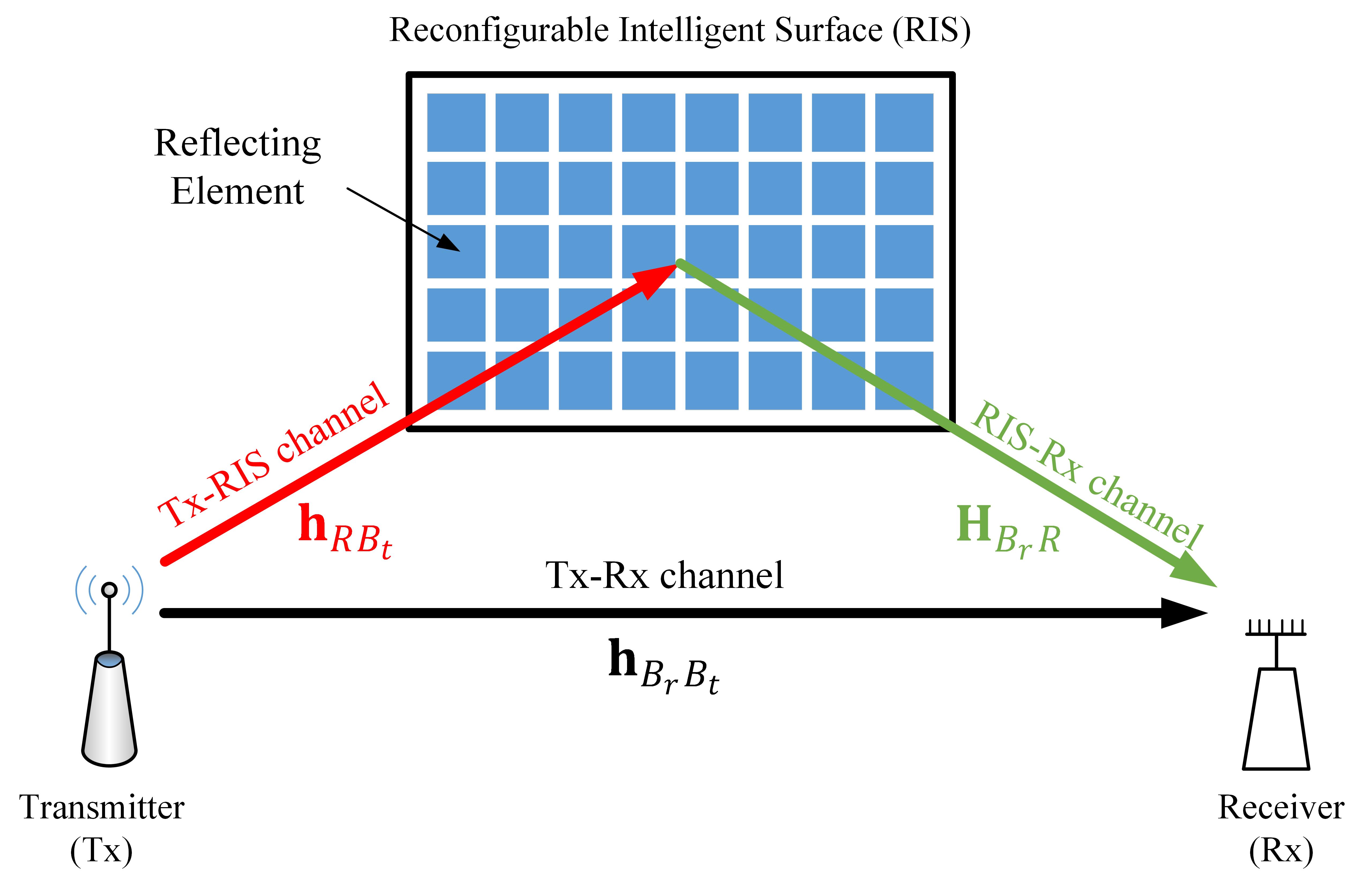,width= 3.in}}
\caption{Signal transmission model for RIS calibration.}
\label{fig.sigModel}
\end{figure}
Consider a single-input multi-output (SIMO) communication system comprised of a single-antenna transmitter, a receiver with $M_r$ antennas, and a RIS with $M_{ris}$ elements as shown in Fig. \ref{fig.sigModel}, where $\hbf_{B_rB_t}\in{\mathbb C}^{M_r\times 1}$ is the channel from the transmitter to the receiver,  $\hbf_{RB_t}\in{\mathbb C}^{M_{ris}\times 1}$ denotes the channel from the transmitter to the RIS, and $\Hbf_{B_rR}\in{\mathbb C}^{M_r\times M_{ris}}$ represents the channel from RIS to the receiver. Given that the transmitter sends pilot sequence $\sbf \in {\mathbb C}^{N\times 1}$ and the RIS is turned on, the receiver receives
\ben
\Ybf = (\Hbf_{B_rR}\Dbf\hbf_{RB_t}+\hbf_{B_rB_t})\sbf^H + \Zbf_{\rm on},
\label{equ.Yon}
\een
where $\Dbf = \diag\left(e^{j\phi_1},e^{j\phi_2},\dots,e^{j\phi_{M_{ris}}}\right)\in{\mathbb C}^{M_{ris}\times M_{ris}}$ and $\{\phi_m\}_{m=1}^{M_{ris}}\in\left[0,2\pi\right]$; $|\sbf(n)|=1,n=1,2,\dots,N$ and $\Enum\{\sbf\sbf^H\}=\Ibf$. We can also turn off the RIS to remove the reflection path of RIS, and then the signal at the receiver becomes
\ben
\Ybf = \hbf_{B_rB_t}\sbf^H + \Zbf_{\rm off},
\label{equ.Yoff}
\een
where $\Zbf_{\rm on}, \Zbf_{\rm off}\in{\mathbb C}^{M_{r}\times N}$ are complex Gaussian noise whose entries follows ${\cal CN}(0,\sigma^2)$; thus, the signal-to-noise-ratio is defined as ${\sf SNR} \triangleq \frac{1}{\sigma^2}$.

\subsection{Phase Deviations}
For $b$-bit RIS, the nominal phase set of its elements is $\{\phi|\phi = (l-1)\Delta,l=1,2,\dots,L\}$ where $L = 2^b$ and $\Delta = \frac{2\pi}{L}$.
In practice, however, the elements of the $b$-bit RIS have different phase deviations across the $L$ gears due to manufactural imperfection and aging, and the phase shift of the $m$-th element at the $l$-th gear is \cite{10023531}
\ben\label{phserr}
\phi_{m,l} = (l-1)\Delta+\epsilon_{m,l},
\een
where $\epsilon_{m,l}$ denotes the phase deviations of the $m$-th element at the $l$-th gear. Thus the phase set of the $M_{ris}$ elements can be denoted as a matrix
$\Pcal\in{\mathbb C}^{M_{ris}\times L}$, of which the $(m,l)$-th element is $\phi_{m,l}$. This paper aims to estimate $\Pcal$ to calibrate the RIS for further applications.

\subsection{Problem Formulation}
According to (\ref{equ.Yon}), the effective channel is estimated as
\ben
\hat{\hbf}_{\rm on} = \frac{1}{N}\Ybf\sbf = \Hbf_{B_rR}\Dbf\hbf_{RB_t}+\hbf_{B_rB_t} + \frac{1}{N}\Zbf_{\rm on}\sbf,
\een
as the RIS is on.
When the RIS is turned off, we have from (\ref{equ.Yoff}) that
\ben
\hat{\hbf}_{\rm off} = \hbf_{B_rB_t} + \frac{1}{N}\Zbf_{\rm off}\sbf.
\een
We can first estimate $\hbf_{\rm on}$ for $Q$ times by changing the gears of the phase shifts on the RIS for $Q$ times and obtain $Q$ pairs of $\Dbf_q$ and $\hat{\hbf}_{{\rm on}, q},q=1,2,\dots,Q$, and then estimate $\hbf_{\rm off}$ for $Q$ times to obtain $\hat{\hbf}_{{\rm off}, q},q=1,2,\dots,Q$, finally we obtain
\ben
\hat{\hbf}_{q}  = \hat{\hbf}_{{\rm on},q}-\hat{\hbf}_{{\rm off},q} = \Hbf_{B_rR}\Dbf_q\hbf_{RB_t} + \zbf_q, q = 1,2,\dots,Q,
\label{equ.hq}
\een
where $\zbf_q\sim {\cal CN}({\bf 0},\frac{2\sigma^2}{N}\Ibf)$ and the phases of $\Dbf_q,q=1,2,\dots, Q$ are some entries of the phase set $\Pcal$. We need to estimate $\Pcal$ by solving the non-convex problem
\ben
\mathop{\min}_{\Pcal,\Hbf_{B_rR}, \hbf_{RB_t}}\ {\sum_{q=1}^{Q}||\hat{\hbf}_{q}-\Hbf_{B_rR}\Dbf_q\hbf_{RB_t}||^{2}_{2}}.
\label{equ.costfunc}
\een

\section{Algorithm for Estimating the Phases with Deviations}
In this section, we first propose a simple way to select a suitable set of gears of the RIS elements in $Q$ measurements; second, we model the RIS calibration problem into a quasi-neural network (QNN) training problem and adopt the well-known backpropagation (BP) algorithm to obtain the estimate of $\Pcal$.
\subsection{Gear Selection} \label{SEC.gear}
Denoting ${\Hbf_{\rm cas}} = \Hbf_{B_rR}\diag(\hbf_{RB_t})\in{\mathbb C}^{M_r\times M_{ris}}$, we have from (\ref{equ.hq}) that
\ben
\hat{\hbf}_{q} = \Hbf_{\rm cas}{\rm exp}(j\phibf_{q}) + \zbf_q, q = 1,2,\dots,Q,
\label{equ.hqv2}
\een
where $\phibf_q(m) = \Pcal(m,\gbf_q(m))$ and $\gbf_q \in {\mathbb C}^{M_{ris}\times 1}$ represents the gears of $m$ elements of RIS in the $q$-th measurement. According to (\ref{equ.hqv2}), we can reformulate (\ref{equ.costfunc}) into
\ben
\mathop{\min}_{\Pcal,\Hbf}\ {\sum_{q=1}^{Q}||\hat{\hbf}_{q}-\Hbf_{\rm cas}{\rm exp}(j\phibf_{q})||^{2}_{2}}.
\label{equ.costfuncv2}
\een
Then we need to decide $\gbf_q,q=1,2,\dots,Q$ for the $Q$ measurements.
Here we divide the $Q$ measurements into $O$ groups, each consisting of $L$ measurements, i.e., $Q = OL$. In the $o$-th group of measurement, the $L$ selected gears of $m$-th element are
\ben
\breve{\gbf}_{o,m} = \left[\gbf_{(o-1)L+1}(m),\gbf_{(o-1)L+2}(m),\dots,\gbf_{(o-1)L+L}(m)\right]^T,
\een
Letting each gear of the phase shifter appears just once in $\breve{\gbf}_{o,m}$, we have
\ben
\breve{\gbf}_{o,m} = \g{\Pi}_{o,m}\cbf\in {\mathbb C}^{L\times 1},
\label{equ.breveg}
\een
where $\g{\Pi}_{o,m}\in {\mathbb C}^{L\times L}$ is a permutation matrix and $\cbf = [1,2,\dots,L]^T\in {\mathbb C}^{L\times 1}$. According to (\ref{equ.breveg}), we can also obtain the $L$ phases of $m$-th element as
\ben\label{qnn.phiom}
\breve{\phibf}_{o,m} = \g{\Pi}_{o,m}\psibf_{m}\in {\mathbb C}^{L\times 1},
\een
where $\psibf_{m} = \left[\phi_{m,1},\phi_{m,2},\dots,\phi_{m,L}\right]^T$ [cf. (\ref{phserr})].
The phases of the $M_{ris}$ RIS elements in the $o$-th group can be denoted as
\ben\label{breveGo}
\breve{\Phibf}_o = \left[\breve{\phibf}_{o,1},\breve{\phibf}_{o,2},\dots,\breve{\phibf}_{o,M_{ris}}\right]^T\in {\mathbb C}^{M_{ris}\times L}.
\een
Hence we can estimate $LM_{ris}$ phases of the RIS with each phase being measured $O$ times.

\subsection{Quasi-Neural Network Formulation}
\begin{figure*}[htb]
\centering
\psfig{figure=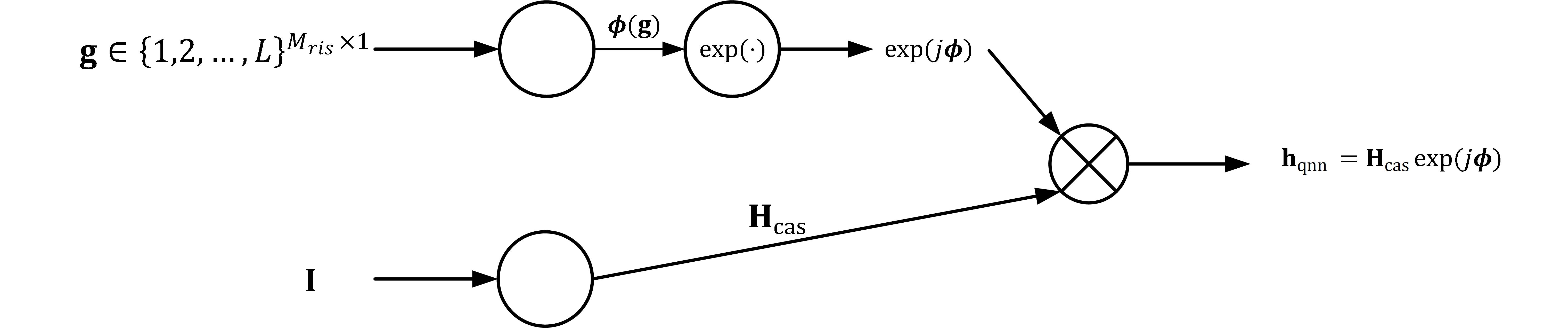,width= 6in}
\caption{The proposed quasi-neural network.}
\label{fig.qnn}
\end{figure*}

We model the RIS calibration problem using a QNN as shown in Fig. \ref{fig.qnn}, which has two inputs: one is the gear vector $\gbf\in \{1,2,\dots, L\}^{M_{ris}\times 1}$; the other is a unit matrix $\Ibf$. The upper branch yields the term ${\rm exp}(j\phibf)$ and the lower branch yields $\Hbf_{\rm cas}$ in Fig. \ref{fig.qnn}, where ${\rm exp}(\cdot)$ acts as an activate function. Feeding $\gbf$ and $\Ibf$ into the QNN, we have the output of the QNN as
\ben
\hbf_{\rm qnn} = \Hbf_{\rm cas}{\rm exp}(j\phibf).
\label{equ.hqnn}
\een
The cost function of the proposed QNN is
\ben\label{qnnCostFunc}
C = ||\hat{\hbf}_q - \hbf_{\rm qnn}||_2^2, q = 1,2,\dots,Q,
\een
where the estimated effective channels $\hat{\hbf}_q,q=1,2,\dots,Q$ can be used as labels to train the QNN.
Using the backpropagation (BP) algorithm, we can obtain the estimated $\Pcal$ as the cost function (\ref{qnnCostFunc})  converges.
\subsection{Backpropagation Algorithm}

The BP algorithm is based on the chain rules, which facilitate the calculation of gradients. To train the proposed QNN, we need to calculate the gradients $\frac{\partial C}{\partial \phibf}$ and $\frac{\partial C}{\partial \Hbf^*_{\rm cas}}$, which is detailed as follows.

First, we calculate $\frac{\partial C}{\partial \phibf}$. We have
\ben
\begin{split}
\frac{\partial C}{\partial \phibf} &=  \frac{\partial C}{\partial \hbf_{\rm qnn}}\frac{\partial \hbf_{\rm qnn}}{\partial \phibf} +
\frac{\partial C}{\partial \hbf_{\rm qnn}^*}\frac{\partial \hbf_{\rm qnn}^*}{\partial \phibf}, \\
&= 2{\rm Re}\left[\frac{\partial C}{\partial \hbf_{\rm qnn}}\frac{\partial \hbf_{\rm qnn}}{\partial \phibf}\right], \\
&= -2{\rm Im}\left[(\Hbf_{\rm cas}^H(\hat{\hbf}_q-\hbf_{\rm qnn}))\circ{\rm exp}(-j\phibf)\right],
\label{equ.Cderphi}
\end{split}
\een
where $\circ$ represents the Hadamard product. Second, we have
\ben
\frac{\partial C}{\partial \Hbf^*_{\rm cas}} = (\hbf_{\rm qnn}-\hat{\hbf}_q){\rm exp}(-j\phibf^T).
\label{equ.CderH}
\een
Then we can update $\Hbf_{\rm cas}$ and $\phibf$ as
\ben
\begin{split}
\Hbf_{\rm cas}(t) &= \Hbf_{\rm cas}(t-1) - \lambda\frac{\partial C}{\partial \Hbf^*_{\rm cas}}(t),\\
\phibf(t) &= \phibf(t-1) - \lambda\frac{\partial C}{\partial{\phibf}}(t),
\end{split}
\label{equ.update}
\een
where $\lambda$ is the learning rate and $t$ is the iteration index for BP algorithm. Iterating $t$ from $1$ to $\infty$ until (\ref{qnnCostFunc}) improves less than $\epsilon$, e.g., $\epsilon = 10^{-5}$, we can obtain the estimated phases $\hat{\Pcal}$.
\subsection{Complexity Analysis}
We conclude the above procedure in Algorithm \ref{Algo.1}, and the computational complexity mainly lies in (\ref{equ.Cderphi}) and (\ref{equ.CderH}) in line $7$, where (\ref{equ.Cderphi}) requires $M_rM_{ris}+\frac{1}{2}M_{ris}$ complex multiplications and (\ref{equ.CderH}) requires $M_rM_{ris}$ complex multiplications.
Hence the complexity of each iteration is $\Ocal(M_rM_{ris})$.

\begin{algorithm}[ht]
\caption{Algorithm for estimating the phases of the RIS}
\label{Algo.1}
\begin{algorithmic}[1]
\Require $Q$ gear vectors $\gbf_{q},q=1,2,\dots,Q$; the effective channel $\hat{\hbf}_{q}$;
\Ensure The estimated phase set $\hat{\mathcal P}$;
\State Initialize the quasi-NN with random $\Hbf_{\rm cas}(0)$ and the nominal phase set ${\mathcal P}(0)$;
\State $t = 1$;
\Do
\State ${\rm loss} = 0$
\For{$q$ = 1 : $Q$}
\State Calculate $\hbf_{\rm qnn}(t)$ by (\ref{equ.hqnn}) based on $\gbf_{q}$;
\State Calculate $\frac{\partial C}{\partial \phibf}(t)$, $\frac{\partial C}{\partial \Hbf^*}(t)$ using (\ref{equ.Cderphi}) and  (\ref{equ.CderH});
\State Update $\phibf(t)$ and $\Hbf(t)$ using (\ref{equ.update});
\State ${\rm loss}  = {\rm loss} + ||\hat{\hbf}_q-{\hbf_{\rm qnn}}(t)||_2^{2}$;
\State $t = t + 1$;
\EndFor
\State ${C}_{\rm ave}  = {\rm loss} /Q$;
\doWhile{${C}_{\rm ave}$ improves less than $\epsilon$.} 
\end{algorithmic}
\end{algorithm}

\section{Cram\'{e}r Rao Bound}
In this section, we first analyze the phase ambiguity and then derive the CRB for the RIS calibration as a performance benchmark of the proposed algorithm.
\subsection{Phase Ambiguity}
We have from (\ref{equ.hqv2}) that
\ben
\hat{\hbf}_{q} = (\Hbf_{\rm cas}\Tbf)(\Tbf^H{\rm exp}(j\phibf_{q})) + \zbf_q, q = 1,2,\dots,Q,
\label{equ.hqv3}
\een
where $\Tbf = \diag(e^{j\varphi_1},e^{j\varphi_2},\dots,e^{j\varphi_{M_{ris}}})$ and $\{\varphi_{m}\}_{m=1}^{M_{ris}}$ are random phases. To remove the ambiguity brought by $\Tbf$, we let
$\varphi_{m} = \phi_{m,1}, m = 1,2,\dots, M_{ris}$,
based on which we derive the CRBs in the next subsection.
According to (\ref{equ.hqv3}), the number of unknown variables is $(2M_r+L-1)M_{ris}$, while the number of constraints is $2LOM_r$. Thus we have
\ben
2LOM_r > (2M_r+L-1)M_{ris},
\een
from which the lower bound of $Q$ is
\ben
Q = OL \ge \left\lceil M_{ris}+\frac{1}{2}(L-1)\frac{M_{ris}}{M_r} \right\rceil.
\label{equ.Obound}
\een

\subsection{Cram\'{e}r Rao Bound}
As the $O$ groups of the measurements are independent, let
us first focus on the $o$-th group to derive the CRB of the
estimations related to the RIS calibration, and the unknown
parameters are the vector
\ben
\etabf = \left[\Omegabf^T, {\rm Re}[{\hbf}_{\rm cas}]^T,{\rm Im}[{\hbf}_{\rm cas}]^T\right]^T\in{\mathbb C}^{M_{ris}(L-1)+2M_rM_{ris}},
\een
where ${\hbf}_{\rm cas} = {\rm vec}\left({\Hbf}_{\rm cas}\right)$ and
\ben
\Omegabf = \left[\phi_{1,2},\phi_{1,3},\dots, \phi_{1,L},\dots, \phi_{M_{ris},2},\dots, \phi_{M_{ris},L}\right]^T.
\een

Combining (\ref{equ.hqv2}) and (\ref{breveGo}) yields that
\ben\label{breveHo}
\breve{\Hbf}_o^T = {\rm exp}(j\breve{\Phibf}_o^T)\Hbf^T_{\rm cas} + \Zbf_o^T
\een
where
\ben
\begin{split}
\breve{\Hbf}_o &= [\hat{\hbf}_{(o-1)L+1},\hat{\hbf}_{(o-1)L+2},\dots,\hat{\hbf}_{(o-1)L+L}], \\
\breve{\Zbf}_o &= [\zbf_{(o-1)L+1},\zbf_{(o-1)L+2},\dots,\zbf_{(o-1)L+L}].
\end{split}
\een
Using formula ${\rm vec}(\Abf\Bbf\Cbf) = (\Cbf^T\otimes \Abf){\rm vec}(\Bbf)$, we have from (\ref{breveHo}) that
\ben\label{breveho}
\breve{\hbf}_{o} = (\Hbf_{\rm cas} \otimes \Ibf_{L}){\rm exp}(j{\rm vec}(\breve{\Phibf}_o^T)) + \breve{\zbf}_o
\een
where $\breve{\hbf}_{o} = {\rm vec}(\breve{\Hbf}_o^T)$ and $\breve{\zbf}_o = {\rm vec}(\breve{\Zbf}_o^T)$. According to (\ref{qnn.phiom}) and (\ref{breveGo}), we can further obtain
\ben\label{qnn.vecbrePhi}
\begin{split}
{\rm vec}(\breve{\Phibf}_o^T) &= \left[\breve{\phibf}_{o,1}^T,\breve{\phibf}_{o,2}^T,\dots,\breve{\phibf}_{o,M_{ris}}^T\right]^T\in {\mathbb C}^{LM_{ris}\times 1} \\
& = \g{\Pi}_o\psibf,
\end{split}
\een
where
\ben
\begin{split}
\g{\Pi}_o &= \diag(\g{\Pi}_{o,1},\g{\Pi}_{o,2},\dots,\g{\Pi}_{o,M_{ris}}), \\
\psibf &= \left[\psibf_{1}^T,\psibf_{2}^T,\dots,\psibf_{M_{ris}}^T\right]^T.
\end{split}
\een
Hence inserting (\ref{qnn.vecbrePhi}) into (\ref{breveho}) yields that
\ben
\breve{\hbf}_{o} = (\Hbf_{\rm cas} \otimes \Ibf_{L})\g{\Pi}_o\psibf + \breve{\zbf}_o
\een
where $\breve{\hbf}_{o}$ is of $\breve{\hbf}_{o} \sim  \mathcal{CN}(\mubf_o, \sigma_z^2\Ibf)$
and $\mubf_o = ({\Hbf}_{\rm cas}\otimes \Ibf)\g{\Pi}_o{\psibf}$. The Fisher Information Matrix (FIM) is
\ben
\Fbf = \frac{N}{\sigma^2}\sum_{o=1}^O{\rm Re}\left[\frac{\partial\mubf^H_o(\etabf)}{\partial\etabf}\frac{\partial\mubf_o(\etabf)}{\partial\etabf}\right].
\label{equ.fisher}
\een
As
\ben
\frac{\partial\mubf_o(\etabf)}{\partial\etabf} = \left[\frac{\partial\mubf_o(\etabf)}{\partial\Omegabf},\frac{\partial\mubf_o(\etabf)}{\partial{\rm Re}({\hbf})}, \frac{\partial\mubf_o(\etabf)}{\partial{\rm Im}({\hbf})}\right],
\label{equ.muodereta}
\een
we can compute $\frac{\partial\mubf_o(\etabf)}{\partial\etabf}$ in three parts. First, we compute $\frac{\partial\mubf_o(\etabf)}{\partial\Omegabf}$. Denoting $\Jbf_o = j({\Hbf}_{\rm cas}\otimes \Ibf)\g{\Pi}_o\diag({\psibf})$, we have
\ben
\frac{\partial\mubf_o(\etabf)}{\partial\Omegabf} = \Jbf_o(:,\Ical),
\label{muoderOmg}
\een
where $\Ical = \{i|i = (m-1)L+(2:L),m=1,2,\dots, M_{ris}\}$.
Second, we have
\ben
\frac{\partial\mubf_o(\etabf)}{\partial{\rm Re}({\hbf}_{\rm cas})} = \left[\rbf_{1,1},\rbf_{2,1},\dots,\rbf_{M_r,1},\dots,\rbf_{M_r,M_{ris}},\right],
\label{equ.muoderReh}
\een
where
\begin{align}
\rbf_{i,j}&=
\left[\begin{array}{l}
\left.
\begin{array}{l}
{\bf 0}_{L\times 1} \\
\vdots  \\
{\bf 0}_{L\times 1}
\end{array}\right\}i-1 \\
\g{\Pi}_{o,j}\psibf_j \\
\left.
\begin{array}{l}
{\bf 0}_{L\times 1} \\
\vdots \\
{\bf 0}_{L\times 1}
\end{array}\right\}M_r-i
\end{array}
\right],
\end{align}
for $i = 1,2,\dots,M_r$ and $j = 1,2,\dots,M_{ris}$.
Third, we have
\ben
\frac{\partial\mubf_o(\etabf)}{\partial{\rm Im}({\hbf}_{\rm cas})} =j \frac{\partial\mubf_o(\etabf)}{\partial{\rm Re}({\hbf}_{\rm cas})}.
\label{equ.muoderImh}
\een
Then inserting (\ref{muoderOmg}), (\ref{equ.muoderReh}), and (\ref{equ.muoderImh}) into (\ref{equ.muodereta}), we have $\frac{\partial\mubf_o(\etabf)}{\partial\etabf}$, from which we can obtain the FIM from (\ref{equ.fisher}). Thus, the CRB is $\Cbf_{\etabf} = \Fbf^{-1}$,
of which the first $M_{ris}(L-1)$ diagonal elements is the CRBs of $\Omegabf$.

\section{Numerical Examples} \label{SEC4}
In this section, we provide some simulation results to verify the effectiveness of the proposed algorithm. The pilot length is 100, i.e., $N = 100$; the permutation matrices are randomly generated; we consider a $4$-bit RIS whose phase deviations are uniformly distributed between $[-20^\circ, 20^\circ]$ \cite[Fig. 1]{9810495}; the learning rate $\lambda$ is set to be $5\times 10^{-3}$. The entries of $\hbf_{B_rB_t}$, $\hbf_{RB_t}$, and $\Hbf_{B_rR}$ are obtained from the Saleh-Valenzuela model \cite{7306370}.

\begin{figure}[htb]
\centering
\psfig{figure=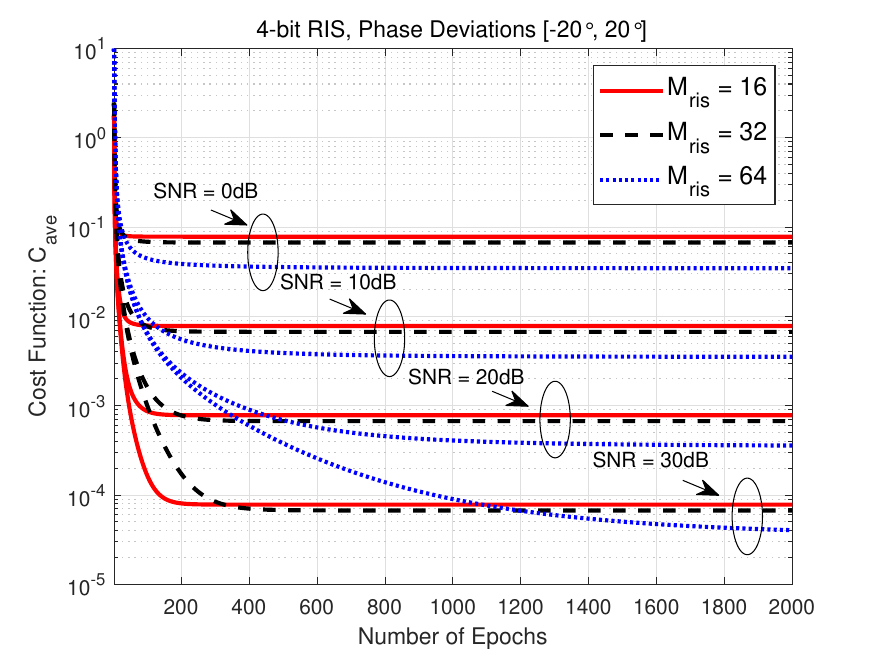,width= 3.in}
\caption{The convergence performance of the proposed algorithm.}
\label{fig.converge}
\end{figure}

In the first example, we simulate the convergence performance of the proposed algorithm using RIS with size of $2\times 8, 4\times 8$, and $8\times 8$ in Fig. \ref{fig.converge}, where the QNN is trained with all $Q$ samples of the training set once in one epoch. Each phase of the $4$-bit RIS is measured $O=15$ times, thus the training set contains $Q = 15\times 2^4 = 240$ samples. The number of
antennas at the receiver is $M_r = 4$. Fig. \ref{fig.converge} shows that the cost function value $C_{\rm ave}$ can achieve a smaller value as SNR varies from $0$dB to $30$dB, where $C_{\rm ave}$ is defined in line $12$ of Algorithm \ref{Algo.1}. Similar to the conventional neural network, the algorithm with RIS of larger size, i.e, $8\times 8$ RIS, tends to converge much slower than that with RIS size of $2\times 8$ and $4\times 8$ as more weights need to be trained.

\begin{figure}[htb]
\centering
\psfig{figure=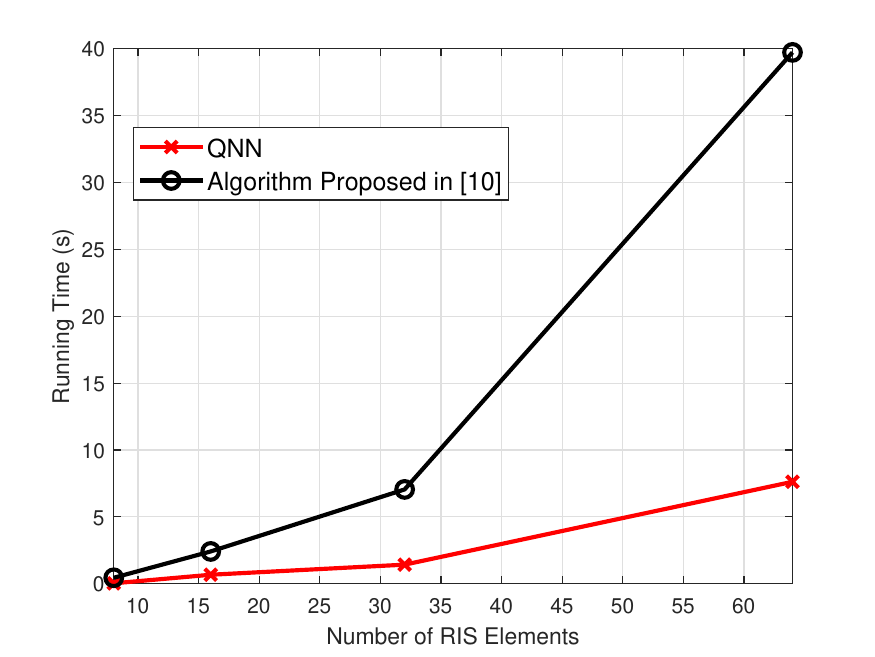,width= 3.in}
\caption{RMSE and running time vs. $M_{ris}$ with $O=15$ and ${\sf SNR} = 20$dB.}
\label{fig.runningTime}
\end{figure}

The second simulation studies the MATLAB (MATLAB R2020b) running time of the proposed BP algorithm, and compares it with that of the algorithm proposed in \cite{10023531} given that where $M_r = 8$. The simulation is conducted on a Windows 10 computer equipped with AMD Ryzen 7 Pro 4750G. According to Fig. \ref{fig.runningTime}, the proposed algorithm takes much less MATLAB running time. That's because the number of optimized variables and the computational complexity in each iteration of the proposed BP algorithm is $M_{ris}$ and $\Ocal(M_rM_{ris})$ rather than $LM_{ris}$ and $\Ocal(M_{ris}^3)$ in \cite{10023531}.

\begin{figure}[htb]
\centering
\psfig{figure=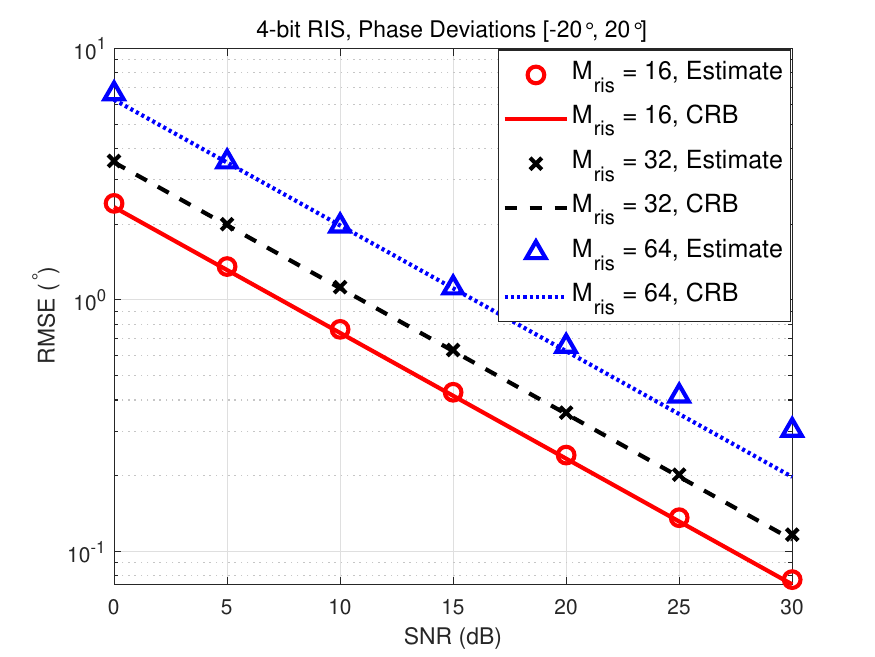,width= 3.in}
\caption{RMSE and CRB vs. SNR with $O=15$.}
\label{fig.snrvary}
\end{figure}

The third example studies the RMSE performance of the proposed algorithm as the SNR varies from $0$dB to $30$dB, where the RMSE is defined as
$$\text{RMSE} \triangleq  \sqrt{\frac{{\mathbb E}\{||\hat{\Omegabf}-{\Omegabf}||_2^2\}}{(L-1)M_{ris}}} \times \frac{180^{\circ}}{\pi}.$$
Fig. \ref{fig.snrvary} shares the same simulation settings with Fig. \ref{fig.converge} and
shows the RMSEs of the estimated phases are close to the CRBs with $M_{ris} = 16,32,64$. But for RIS of larger size, e.g., $8\times 8$ RIS, the proposed algorithm tends to converge at a locally optimal point at SNR $= 30$dB, and hence the RMSE slightly deviates from the CRB.

\section{Conclusion} \label{SEC5}
This paper proposes to model the estimation of the RIS phase with deviations into a quasi-neural network (QNN) training problem and utilize the backpropagation (BP) algorithm to train the QNN, which can estimate the phases of the RIS elements at each gear given unknown channel state information (CSI). We also derive the Cram\'{e}r Rao Bounds (CRBs) for the RIS calibration as a benchmark of the proposed approach. The simulation results verify the effectiveness of the proposed algorithm as the root mean square errors (RMSEs) of the phase estimates are close to the Cram\'{e}r Rao Bounds (CRBs), and show its superior performance in computation complexity over the state-of-the-art technique.

\bibliographystyle{IEEEtran}
\bibliography{bib}
\end{document}